# SFX Miscellaneous Free Ejournals Target: Usage Survey Among the SFX Community

François Renaville

*University of Liège Library, University of Liège, Liège, Belgium*

Yosef Branse

*Younes and Soraya Nazarian Library, University of Haifa, Haifa, Israel*

Xiaotian Chen

*Cullom-Davis Library, Bradley University, Peoria, Illinois, USA*

Mark Needleman

*Library Services Division, Florida Virtual Campus, Gainesville, Florida, USA*

**ABSTRACT**

The number of free or open access articles is increasing rapidly, and their retrieval with library indexes and OpenURL link resolvers has been a challenge. In June 2014, the SFX MISCELLANEOUS_FREE_EJOURNALS target contained more than 24,000 portfolios of all kinds. The SFX KnowledgeBase Advisory Board (KBAB) carried out an international survey to get an overview of the usage of this target by the SFX community and to precisely identify what could be done to improve it. The target is widely used among the community. However, many respondents complained about three major problems: (a) incorrect links, (b) full texts actually not free, and (c) incorrect or missing thresholds (years and volumes information).

*******************************************

## INTRODUCTION

SFX was the first OpenURL link resolver and remains the one most widely used, having been adopted by over 2,400 libraries (Ex Libris SFX, n.d.b). The OpenURL framework has its roots in the SFX research conducted by Herbert Van de Sompel, Patrick Hochstenbach, and colleagues at Ghent University (Belgium) from 1998 to 2000 (Van de Sompel and Beit-Arie, 2001). In early 2000, Ex Libris Group acquired the SFX server software from Ghent University. Ex Libris reengineered the software and marketed it to libraries as an autonomous, reference linking service component that fits in the OpenURL framework.

The OpenURL framework was enhanced by the NISO (National Information Standards Organization) OpenURL standard (NISO, 2005) that defines the OpenURL version 1.0. The NISO OpenURL standard was further standardized and developed for defining metadata element sets so that the OpenURL could be used in a wide variety of domains beyond what had been defined by the researchers at Ghent University in their original document. That version of the OpenURL is now referred to as OpenURL 0.1. SFX supports both OpenURL 0.1 and the OpenURL 1.0 NISO standard.

The strength of SFX rests also on a huge "Knowledge-Base" (henceforth: KB), which "serves as a central data repository to support all SFX features and services and to furnish institutional availability information to external systems. Information from the SFX KB is the key to linking end users quickly and accurately to the material they need" (Ex Libris SFX, n.d.a).

The quality of the SFX KB relies on the interaction of three agents:
(1) Ex Libris KB team collects, checks, enriches, and corrects data on behalf of the customers.
(2) The publishers/vendors/aggregators provide the necessary materials to increase and improve the content of SFX KB.
(3) The libraries whose staff (mostly local SFX administrators) participate in the KB development by suggesting new content and reporting to the KB team errors such as broken links and coverage problems.

Librarians' contributions can be reported to Ex Libris via the customer support portal (Salesforce), the "Send to Ex Libris" link available for SFX customers via the SFX Administrator (Admin) interface, or, since spring 2013, via the "Contribute" button, also directly available in the SFX Admin. The Contribute button enables libraries to actively share local data with the global SFX community. Libraries can disseminate local changes directly from the SFX Admin by clicking the new Contribute button (visible in the "View Object Portfolio" window) for the "Miscellaneous" targets, notably the MISCELLANEOUS_FREE_EJOURNALS (henceforth: MFE) target that contains more than 24,000 object portfolios. In SFX terminology, a *target* is a user's selected resource, often but not always full text, or a link presented on the SFX menu that allows the user to get content and services; an *object portfolio* is a journal or book within a target with specific years or volume information.

Maintaining the quality of a database as large as the SFX KB is a very difficult, if not endless, task in terms of quality of the metadata and of linking parameters such as thresholds, parse params, etc. In SFX terminology, a *threshold* contains information related to dates, volumes, and issues of journals available to users. *Threshold* is a PERL statement that delineates access conditions for an object portfolio. These conditions typically consist of publication information such as year published, volume, and issue (a parsed Date threshold). The threshold can also contain other information such as embargos and ending dates for access. A *parse param* contains information that is used by the target parser to create URLs. The parse param can also contain variables for data such as user names and passwords that differ from one institution to the next.

Due to the challenges of maintaining SFX KB, the International Group of Ex Libris Users (IGeLU) and the Ex Libris Users of North America (ELUNA) decided to create a joint group whose goal is to promote first-class quality of the data stored in the SFX KB by reviewing the quality assurance policies and processes together with Ex Libris. The SFX Knowledge Base Advisory Board (KBAB) was founded in 2013 as a result of discussions at the IGeLU 2012 Conference in Zurich.

Free or open access (OA) journals grew rapidly in numbers and prominence during the period that OpenURL technology and SFX were developed. On April 1, 2009, the list of OA journals included in the Directory of Open Access Journals (DOAJ) reached the mark of 4,000 journals, and by December 31, 2014, the number of DOAJ journals had more than doubled to 10,134 (DOAJ, 2014). In the "Gold

OA" model, all the articles of a certain journal are freely accessible online without any restriction or embargo (Hamad et al., 2004). Lewis (2012) projected that "Gold OA could account for 50% of the scholarly journal articles sometime between 2017 and 2021, and 90% of articles as soon as 2020 and more conservatively by 2025."

With these perspectives, KBAB decided in 2014 to find out how free journals are being handled in SFX.

**LITERATURE REVIEW**

OpenURL link resolvers have never been 100% accurate. Wakimoto, Walker, and Dabbour (2006) estimated that the overall SFX failure rate—caused by a variety of reasons—is about 20%, and they divided errors into two broad categories: "either because they incorrectly showed availability (false positives) or incorrectly did not show availability (false negatives)." Users generally do not notice false negatives, since they fail to show article availability. False positives are dead links users see. Trainor and Price (2010) not only confirmed the Wakimoto et al. findings but found that SFX full-text linking accuracy did not improve from 2004 to 2010. The error rates of other OpenURL link resolvers are about the same. When comparing two other Open URL link resolver products (360 Link and a local product), Herrera found that the error rate of OpenURL linking is comparable to what Wakimoto et al. found six years earlier: 16% of user-generated OpenURLs had technical issues; some had more than one (Herrera, 2011). Chen (2004), Donlan (2007), and Wakimoto et al. (2006) all found that wrong information provided by database vendors is a major cause of errors. McCracken and Womack summarized OpenURL linking problems this way: "bad data from the provider, bad formatting of the data, and a lack of knowledge among the members of the data supply chain" (McCracken and Womack, 2010). In addition to commonly discussed dead link causes, such as incorrect metadata, resolver translation error, inaccurate embargo data, provider target URL translation error, incomplete provider content, wrong coverage dates, and indexed-only titles mistakenly considered as full-text titles, Chen (2012) also called for attention to other causes of dead links: book reviews and other special items that do not work with OpenURL linking; articles from supplemental issues that have nonstandard metadata, abnormal volume, issue and page numbers; articles with two publication dates (one traditional, the other online); DOI errors; and missing volumes/issues/articles on journal websites. Classer described three industry initiatives (Knowledge Bases and Related Tools [KBART], Improving OpenURLs Through Analytics [IOTA], and Presentation and Identification of E-Journals [PIE-J]) aimed at improving access to licensed electronic content, with KBART and IOTA focusing on "metadata inaccuracies that affect the efficacy of OpenURL linking," and PIE-J "working toward the creation of a NISO-recommended practice that will guide providers on how best to present journal title and ISSN information" (Classer, 2012). In discussing cataloging OA journals, Schmidt and Newsome (2007) touched on the difficulties in maintaining OA journals in their online public access catalog. It is clear that the existing literature on OpenURL linking errors focuses on electronic journals with a traditional subscription model.

It is unknown whether or not Lewis's projection that Gold OA could account for 50% of the scholarly journal articles sometime between 2017 and 2021 will become reality. But Morrison estimated in 2012 that "About 30% of peer reviewed scholarly journals are now open access," based on the DOAJ number and her estimate that "active, peer-reviewed scholarly journals from Ulrich's using deduplication from December 1, 2011 is 26,746" (Morrison, 2012). It is probably safe to say that in 2014, two years after Morrison's 30% estimate, 30% or more peer-reviewed journals are Gold OA or free journals. With OA or free journals growing like this, it is appropriate to study how OpenURL linking works with them.

**WHY CONDUCT A SURVEY ON THE MFE TARGET?**

There are various free or OA journal targets in the SFX KB, such as DOAJ_DIRECTORY_OPEN_ACCESS_ JOURNALS_FREE, PUBMED_CENTRALJOURNALS_ FREE, and TAYLOR_FRANCIS_OPEN_ACCESS_FREE. However, MFE is by far the largest free target, with more than 24,000 journals of all kinds. The DOAJ target, with about 10,000 journals, is the only free target that is close to MFE in size; most of the other free targets range in size from under 100 to several thousand titles. In addition to size, a major difference between MFE and the other free targets is that the rest all have a vendor or owner, such as DOAJ, PubMed Central, and Taylor & Francis, that manages and supplies the data to SFX, while the MFE target is created and maintained by Ex Libris with the help of the clients (via the local SFX administrators). MFE was often mentioned in messages sent to the SFX international discussion list SFX-DISCUSS-L at the end of 2013 and early 2014. SFX administrators pointed out then some weaknesses of that catch-all target and complained about many errors (e.g., broken links, incorrect thresholds, not all contents are free). In April 2014, KBAB members decided to analyze a random sample of the existing 24,236 object portfolios and systematically checked all parse params of 484 object portfolios (i.e., every 50th), that is to say 2% of the whole target. They discovered that one out of five parse params of the sample was not correct (Renaville & Needleman, 2014). This 20% rate of incorrect parse params is significantly higher than that of other targets. It was mentioned in the literature review section that Wakimoto et al. estimated in 2006 that the overall SFX error rate is about 20% (Wakimoto et al, 2006). The 20% rate includes false positives (dead links) and false negatives (no full-text links offered when full texts available), with a large number of reasons for the occurrence of false positives. On the other hand, one category (incorrect parse params) alone leads to 20% false positives (dead links) for MFE. Besides incorrect parse params, there are numerous other reasons for the occurrence of false positives (dead links), such as resolver translation error, inaccurate embargo data, provider target URL translation error, incomplete provider content, wrong coverage dates, indexed-only titles mistakenly considered as full-text titles, and other reasons listed in the literature review section. Because MFE has a higher percentage of dead links and is the largest free journal target on SFX, it was chosen by KBAB for further study.

In order to get a representative idea of the usage that is done by librarians of the MFE target and to precisely identify what could be done to improve it, KBAB decided to launch a survey among the SFX community. The survey, containing eight multiple-choice and open questions related to that target, was posted to the listserv SFX-DISCUSS-L in June 2014 and was open from June 23 until July 16, 2014. SFX-DISCUSS-L has about 1,400 subscribers, who are typically local SFX administrators. As it was an institutional survey, only one response per institution was expected. The question list is to be found in the appendix.

## SURVEY RESULTS AND ANALYSIS

The survey received 122 responses from 18 countries: Australia (5), Belgium (3), Brazil (2), Canada (3), Denmark (2), France (7), Germany (6), Israel (4), Italy (2), the Netherlands (3), New Zealand (1), Norway (2), Portugal (1), Spain (2), Sweden (3), Switzerland (3), United Kingdom (11), and United States (62).

### Question 1: Do you Use the MISCELLANEOUS_FREE_EJOURNALS Target?

Of the 122 respondents, 114 (93.4%) use the target; only 8 (6.6%) do not. The reasons put forward for not using it are linked to the quality of the database and to the error rate, which appeared to be very high:

- Broken links (2).
- Incorrect thresholds (3).
- Journals offered articles for purchase (2).
- Many journals are not scholarly journals, have a lot of advertisements (2).
- Quality of the objects' metadata (languages, publishers, ISSN) (1).

Moreover, one respondent explained they have hardly enough "staff or time to fully maintain the resources that [they] actually pay for" and that free electronic resources, especially if their quality is not appropriate, have less priority. Another respondent assured they would use the MFE target:

- If local updating of portfolios could more systematically be used to contribute to improve the Central KB quality (for example with a pop-up invitation to use the Contribute button every time thresholds or parse params have been locally updated).
- If languages, publishers, and ISSN could be updated locally.
- If portfolios (especially thresholds and parse params) were more systematically checked.
- If scholarly journals could be set apart from nonacademic journals.

Two respondents acknowledged that they previously activated that target but deactivated it later for quality reasons and because "libraries received complaints from their users."

Those who answered *No* to this first question were brought to Question 8, Questions 2-7 being unavailable to them.

### Question 2: Do You Activate All of the Object Portfolios in the Target or Only Selectively?

"Activate" in SFX terminology is the act of turning on targets or objects in the local SFX Administration site. Of the 114 users of the target, 48 respondents (42.1%) activate all the object portfolios; 66 (57.9%) activate only a selection. The most frequent reasons for *all* activation include:

- Not enough staff to go through the whole target and check each portfolio (19).
- Want to provide as many (free) resources to users as possible (10).
- Rely on Central KB quality and presume that free full-text is available for all titles (4).
- Request by specialist/public services librarians (2).

In eight cases, the respondents explained they activate all portfolios but then deactivate individual titles where problems are reported by users or other librarians. In most comments, respondents specify that the problems are then reported to Ex Libris.

The reasons for selective activation are varied and include:

- Only activate titles that are requested or of interest to the users (36).
- Links are too often broken (18).
- Titles are not really free or Open Access (10.
- Duplicates other preferred targets like DOAJ or HIGH-WIRE_PRESS_FREE (8).
- Incorrect thresholds (4).

Of the 36 respondents who activate portfolios on the basis of what could be of interest to their users, several point out that the target contains a lot of uninteresting titles and low-quality material; three focus on the language variety as a source of annoyance (English-only titles are preferred); one says that the "target is not updated often enough to be reliably accurate," without providing any additional detail.

Most of the reasons that explain why libraries do selective activation are the same, as those that are put forward for not using MFE at all.

**Question 3: Do You Set Auto-Activate on for This Target?**

"Auto-activate" is an option in SFX Administration that turns on new object portfolios automatically when they are added to a target. Of the 114 users of the target, 56 respondents (49.1%) have set auto-activate on for MFE; 58 (50.9%) have not.

Of the 58 who do not use the auto-activate function, 24 (41.4%) point out it is important to them to manually select and check any title available to their end users ("Because we feel that we would have to test each link individually; there is little credibility," "We do not want to add titles we do not know anything about," "Too much noise for 'All portfolios,'" "Very much of them are inappropriate for us").

Seven of them (12.4%) complain that titles are not always free. This is rather surprising, since auto-activate is intended to automatically activate newly added portfolios with the weekly KB updates. One would rather expect that new portfolios added by Ex Libris are correct and provide accurate information that has normally been checked very recently. Therefore, parse params, thresholds, and the free nature of the new portfolios are expected to match the reality. This complaint that titles are not always free in MFE, even in the case of newly added portfolios, indicates a measure of distrust of KB updates.

**Question 4: Do You Add Local Object Portfolios to This Target?**

SFX allows local administrators to add to a target local object portfolios of ejournals and other items that were not centrally or globally included in a target from the KB. Of the 114 users of the target, 36 respondents (31.6%) say they never add any local object portfolios to the MFE target, 28 (24.6%) rarely do, 38 (33.3%) sometimes add some portfolios, and 12 (10.5%) often do (see Figure 1).

The 50 respondents who answered "Yes" were asked a complementary question related to their involvement in the growth of the KB.

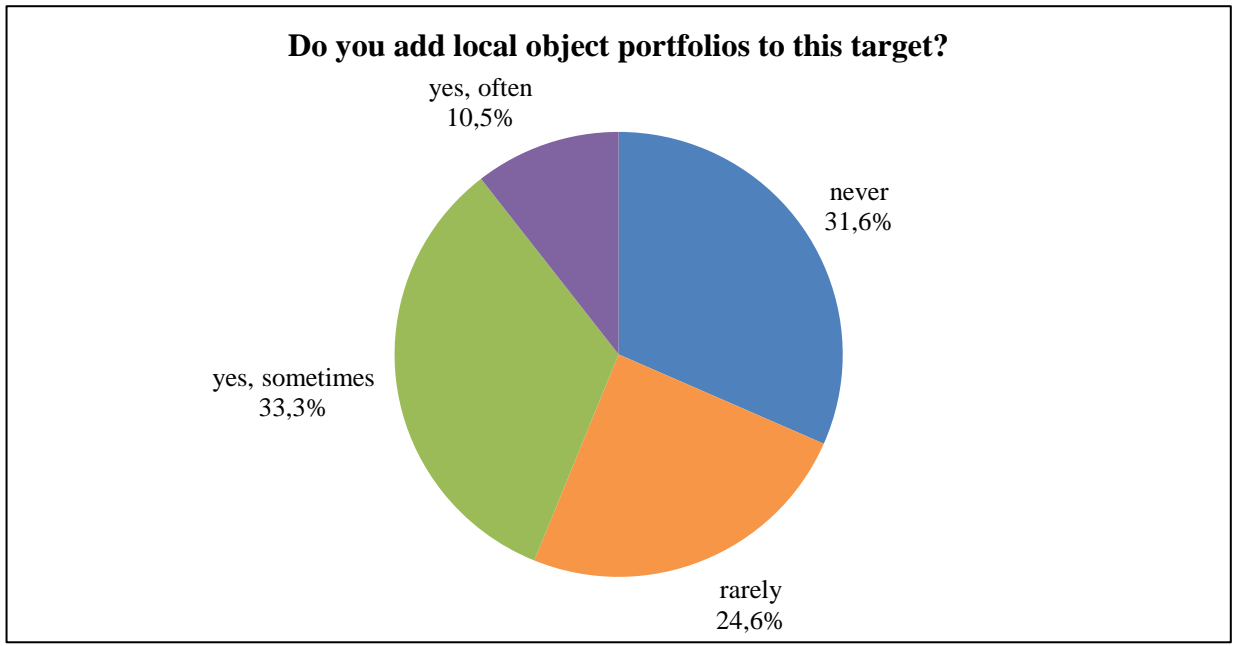

*Figure 1. Adding local objects to MFE target.*

**If Yes, Do You Contribute? Do You Request Ex Libris to Add Them to the KB?**

Of the 12 who said they often add local portfolios, four respondents (33.3%) said that they never request Ex Libris to add them in the KB, three (25.0%) said they rarely ask for it, two (16.7%) sometimes, and three (25.0%) often. Thus in this survey sample, more than half of those who (very) actively create local portfolios rarely or never request Ex Libris to add them in the KB (see Figure 2).

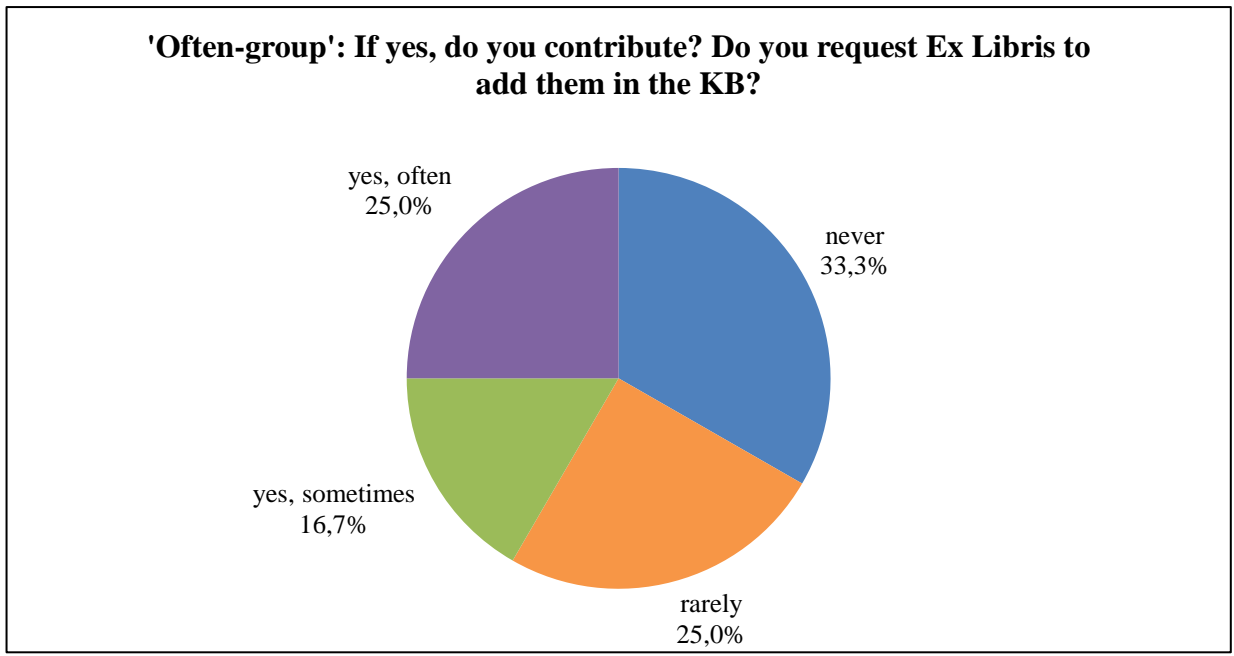

*Figure 2. Often-group: Requesting Ex Libris to add local objects to Global KB.*

Of the 38 who said they sometimes add local portfolios, one respondent (2.6%) said that s/he never requests Ex Libris to add them in the KB, 12 (31.6%) said they rarely ask for it, 12 (31.6%) sometimes, and 13 (34.2%) often. In other words, more than two-thirds of those who sometimes locally create portfolios take the trouble to share the information with Ex Libris and to contribute to the growth of the target in KB (see Figure 3).

From those responses, it appears that those who are the most active in contacting Ex Libris to get local objects added as global objects in the KB are those who also create fewer local portfolios. Those who create more local portfolios, certainly considering it necessary to improve their own KB, are less inclined to contribute to the KB.

The number of responses (50) is certainly too low to draw any clear and definite conclusions, but the findings that the more active one is locally the less active one is globally would certainly need to be examined in more detail.

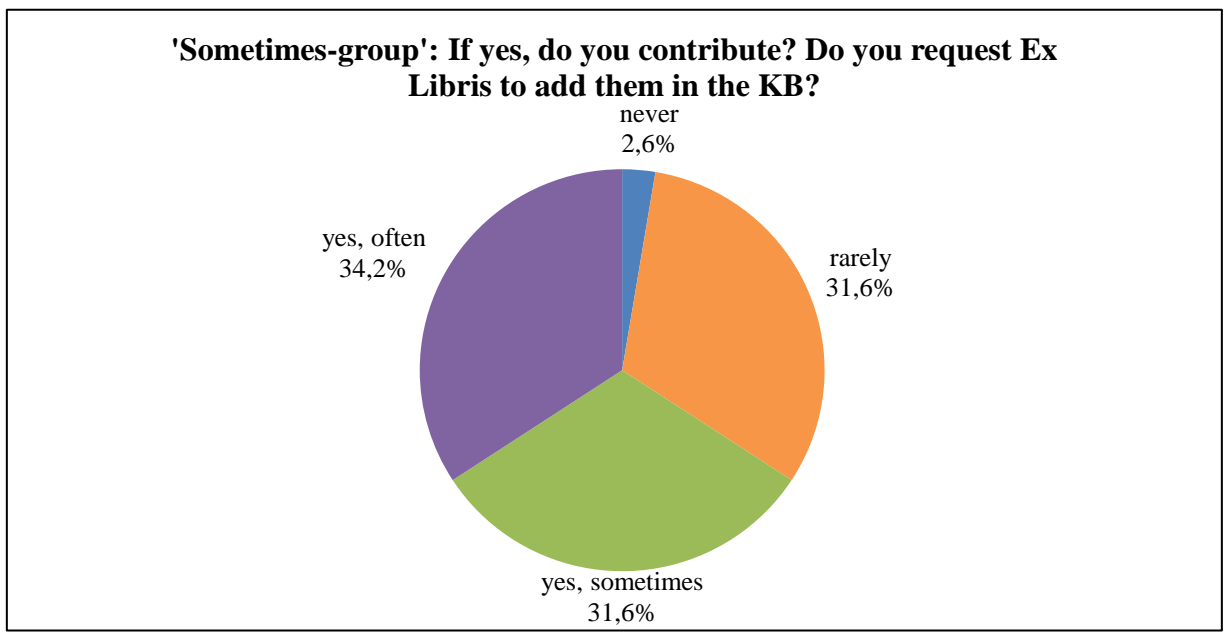

*Figure 3. Sometimes-group: Requesting Ex Libris to add local objects to Global KB.*

## Question 5: Do You Locally Correct Thresholds or Parse Params of Any of the Portfolios in This Target?

Of the 114 users of the target, 11 respondents (9.6%) answered that they never locally correct thresholds or parse params of any of the MFE portfolios, 32 (28.1%) rarely do, 47 (41.2%) sometimes locally correct thresholds or parse params, and 24 (21.1%) often do. So, 103 respondents (90.4%) feel the need to varying degrees to make up for the errors or lack of precision within the MFE target (see Figure 4).

The 71 respondents who answered "Yes, sometimes" or "Yes, often" were asked a complementary question related to their involvement in the quality of the KB.

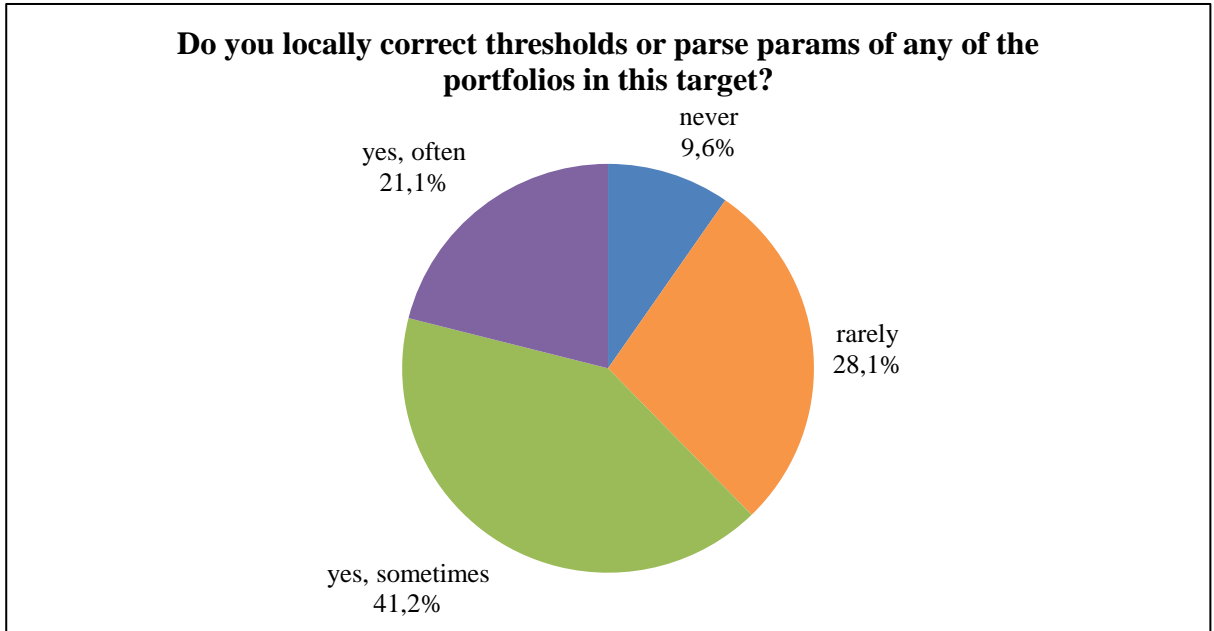

*Figure 4. Locally correcting thresholds and/or parse params.*

**If Yes, Do You Contribute? Do You Request Ex Libris to Correct Them in the KB?**

Of the 24 who said they often locally correct thresholds or parse params within the target, six respondents (25.0%) said that they never request Ex Libris to correct the portfolios, seven (29.2%) said they rarely ask for it, four (16.7%) sometimes do, and seven (29.2%) often do (see Figure 5).

Less than half of those who are very active in locally correcting portfolios take the time to contribute and to inform Ex Libris in order to improve the quality of the KB.

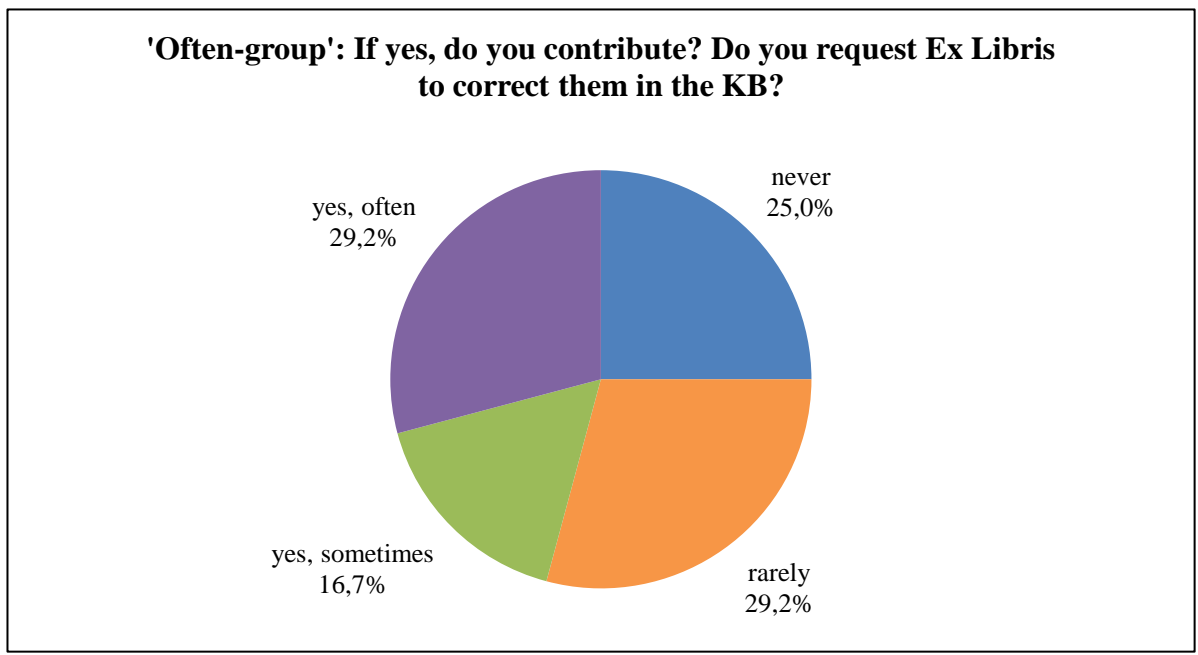

*Figure 5. Often-group: Requesting Ex Libris to correct Gobal KB.*

Of the 47 who said they sometimes locally correct thresholds or parse params, five respondents (10.6%) said that they never request Ex Libris to correct the portfolios, 10 (21.3%) said they rarely ask for it, 21 (44.7%) sometimes do, and 11 (23.4%) often do (see Figure 6).

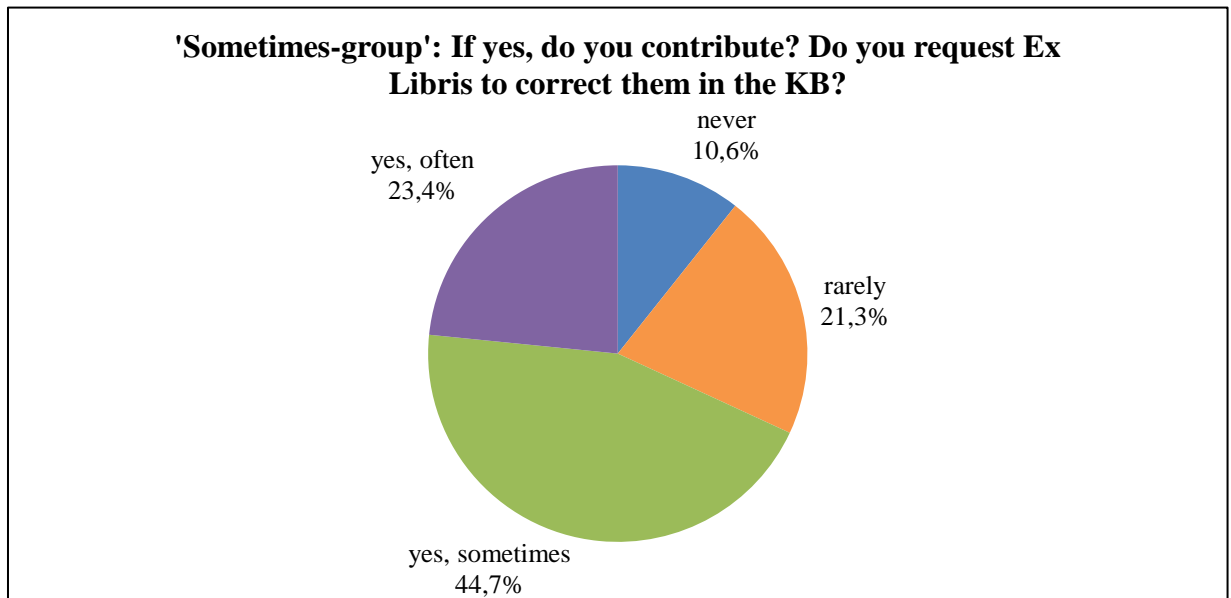

*Figure 6. Sometimes-group: Requesting Ex Libris to correct Global KB.*

Here again, it appears that those who are less active in locally correcting MFE portfolios are much more inclined to contribute and to improve the general quality of the KB.

The 28 respondents who answered "Never" or "Rarely" to the additional question "If yes, do you contribute? Do you request Ex Libris to correct them in the KB?" were also asked to explain why they do not or rarely contribute.

**If Rarely or Never, Please Explain Why**

Not everyone answered this question; however, two main reasons come up:
- 10 respondents pointed out it would take them too much time to contribute.
- 7 admitted they did not know it is possible or do not know how to configure the Contribute functionality.

Finally, one said that it is useless; another said that the work should be performed by the Ex Libris teams. Here are examples of "No Time" feedback:
(1) "It is an extra step to click on another button and open a new window."
(2) "It takes time to report them and there is a good chance that the information will have changed by the time the change is implemented. I also already spend enough time reporting other issues (some related to SFX, some not) to vendors."
(3) "No time to do it."
(4) "I find the process cumbersome to contribute, and sometimes Ex Libris takes a very long time to incorporate. It doesn't seem worthwhile."
(5) "I have not updated/configured my instance to be able to contribute (yet) and in the past asking EL to correct the data in this target generally led to them saying they did not manually correct this target but relied on feeds only and so it was a waste of my time."
(6) "Because corrections seem to never show up in the KB, it is a waste of my time."
(7) "No time."

(8) "Low priority, large amount of work."
(9) "It's just another thing to do; Ex Libris should do this."
(10) "It's another action, logging into Salesforce, raising a query. I already have too many changes on the KB."

Here are examples of "Do Not Know" feedback:
(1) "Because we never did set SFX up to automatically report to Ex Libris."
(2) "The way to do this has never been clearly communicated to me."
(3) "Not sure how it would work, with our consortium set-up. Have had little to no training in use of SFX."
(4) "Send to Ex Libris' button didn't work when last tried."
(5) "I'm quite new to SFX and did not know you could do this."
(6) "Because we didn't realize we could."
(7) "Since the new Salesforce system we can't rapport [sic] directly from our SFX instance—unfortunately. Our IT department hasn't prioritized this task."

From the comments, it appears that Ex Libris should certainly be advised to communicate about the Contribute button and to find ways to make it work more quickly and with as little configuration as possible.

**Question 6: What Would You Like to See Happen With This Target? What Could Ex Libris Do, in Your Opinion, to Improve This Target?**

To answer this question, five statements were proposed to the respondents. They were asked to grade each of them from 1 *(Least Desirable)* to 5 *(Very Desirable),* with 3 being the most neutral position. More precisely, the first statement was related to what Ex Libris could do to improve the target as it is now. The second and third were suggestions related to the future of the target (what could happen to it). The fourth and fifth statements concerned the Contribute button.
(1) Checking the portfolios (thresholds and parse params) more systematically?
(2) Creating a new specific MISCELLANEOUS_FREE_SCHOLARLY_EJOURNALS for academic journals only? All the other journals would then stay in MFE.
(3) Deleting from MFE all portfolios that already exist in another FREE or OPEN-ACCESS target?
(4) Promoting the use of Contribute button?
(5) Getting a pop-up invitation to use the Contribute button every time I have locally updated a threshold or a parse param?

**Checking the Portfolios (Thresholds and Parse Params) More Systematically?**

Of the 114 respondents, 95 (83.3%) think that Ex Libris should pay more attention to checking the portfolios of the target more systematically (see Figure 7). Very few—only seven (6.1%)—consider it to not be a priority. However, some respondents have doubts about how quality control can practically be maintained for more than 24,000 records and suggest a closer collaboration with clients:

> *Checking portfolio thresholds and parse params would be great but unrealistic. This should be done centrally but I do not expect Ex Libris to do it. (Could this be a cooperative venture among volunteer libraries?) A smaller target that is well maintained is a more realistic and*

> *better alternative. Ex Libris already does a good job of promoting use of the Contribute button, catching me when I forget.*

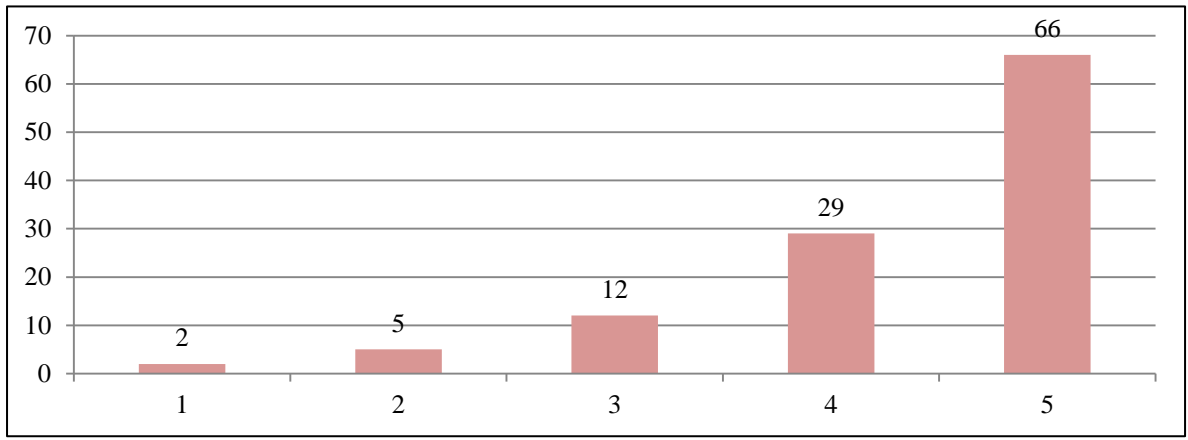

*Figure 7. Checking object parameters more systematically.*

### Creating a New Specific MISCELLA-NEOUS_FREE_SCHOLARLY_EJOURNALS for Academic Journals Only? All the Other Journals Would Then Stay in MFE

Globally, 64 respondents (56.1%) do think it would be a good idea to take scholarly journals out of the MFE and to create a new specific MISCELLA-NEOUS_FREE_SCHOLARLY_EJOURNALS for academic journals only, while 28 (24.6%) do not agree (see Figure 8).

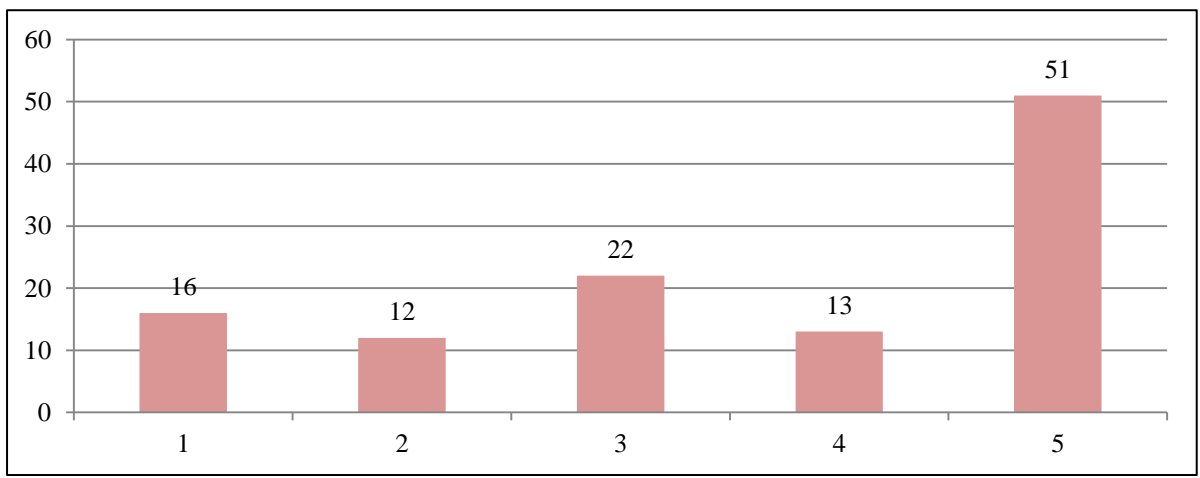

*Figure 8. Creating a new miscellaneous free target for academic journals only.*

However, if such a new target is created, one has to define how the scholarly nature of an object would be established, the peer-reviewed criterion only being inadequate because it's too restrictive, especially for free journals. This would certainly not be an easy task.

Here again, some respondents also point out that Ex Libris would certainly need the collaboration of libraries to create the target, if not thematic subdivisions like MISCELLANEOUS_FREE_SCHOLARLY_EJOURNALS.BUSINESS, MISCELLANEOUS_FREE_SCHOLARLY_EJOURNALS_ MEDICINE, etc.

For another, a smaller, more-focused target may make things easier to maintain: "If Ex Libris wants to prioritize maintenance for a smaller target, then OPEN_ACCESS_SCHOLARLY_EJOURNALS may be a useful target to establish."

Finally, a user finds there is a real opportunity for information literacy in separating scholarly journals from the other ones: "Separating out the scholarly would help me show my students more easily what the differences are between scholarly and non and enable them to be more confident about the choices they make independently."

**Deleting From MFE All Portfolios That Already Exist in Another FREE or OPEN_ACCESS Target?**

A majority of 81 respondents (71.1%) like the idea of "de-duplicating" the target from other existing free or OA targets, so that MFE has only unique entries (see Figure 9).

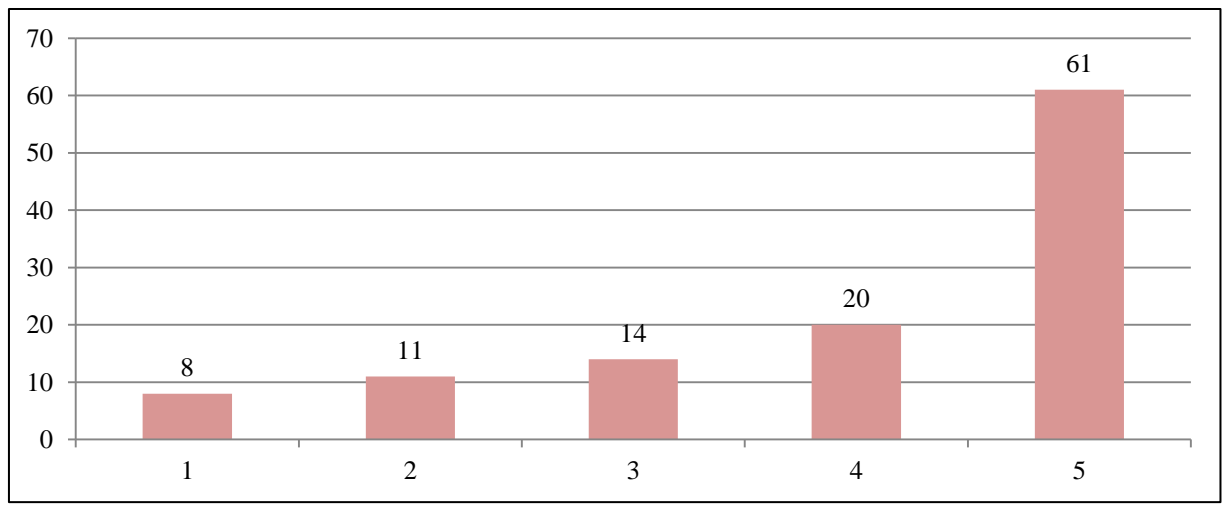

*Figure 9. Deleting objects from MFE target that exist in another free or Open Access target.*

A user suggests that deleting an object portfolio from MFE is controversial because one given institution may have only enabled this target and explains that to hide a portfolio from this target in the SFX menu, one can simply add a new display logic rule like "if any getFullTxt available > do not show MFE." However, using logic rules in order to hide MFE from the SFX menu may also be a drastic solution with very little flexibility, if any.

**Promoting the Use of the Contribute Button?**

Less than half of the respondents (49.1%) think that the use of the Contribute button should be promoted among the SFX community, one-third of the respondents have no real opinion, and 20 (17.5%) are against that idea (see Figure 10).

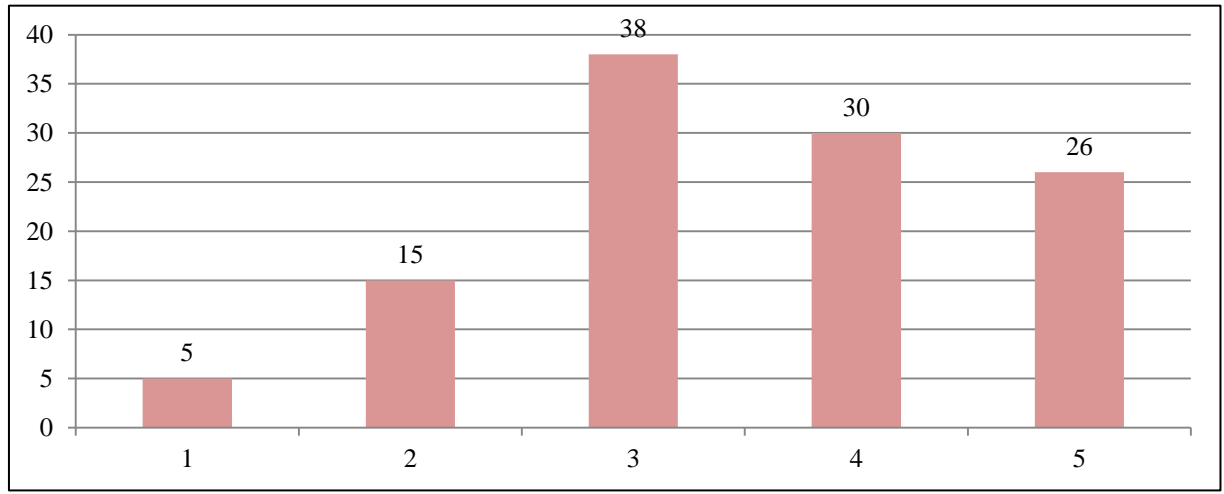

**Figure 10.** *Promoting use of the Contribute button.*

## Getting a Pop-Up Invitation to Use the Contribute Button Every Time I Have Locally Updated a Threshold or a Parse Param?

Only 55 users (48.2%) think that getting a pop-up invitation to use the Contribute button would be an improvement, while 29 (25.4%) are against it. Of the five statements, this is the one with the most people against it (see Figure 11). From the comments, it appears that getting a pop-up is considered too invasive a solution:

- "I think a pop-up is too invasive, but it would be good to make it easier for sites to Contribute."
- "The promotion for the Contribute button is unnerving already. [...]. An additional pop-up would be further unnerving."
- "And, if the Contribute button is somehow promoted (NOT with a pop-up, because those interfere with workflow), that would be a great community asset."

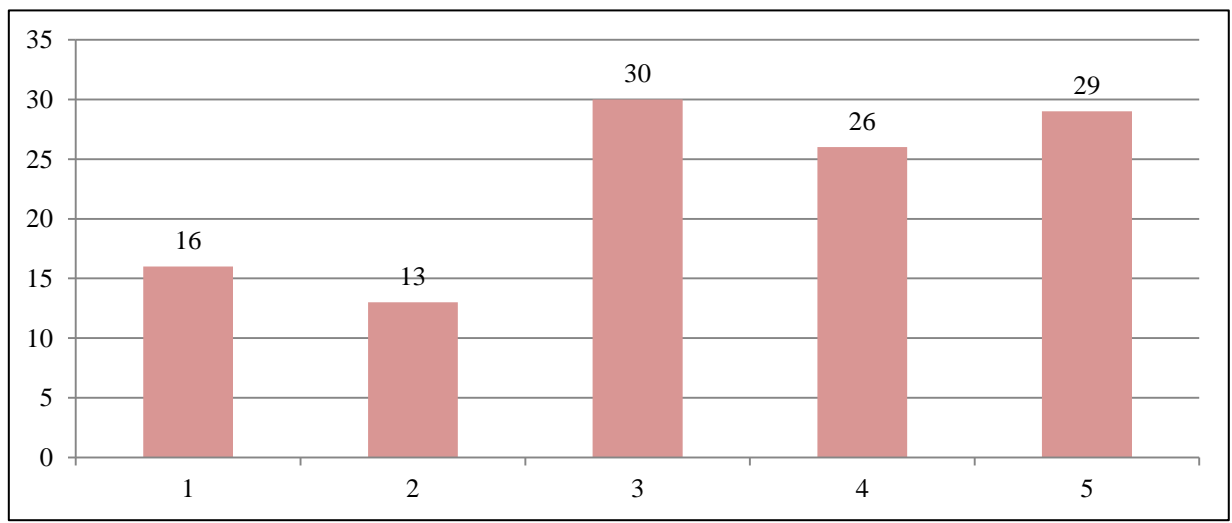

*Figure 11. Getting pop-up invitation to contribute locally added object parameters.*

The high percentage of indecisive users for questions 6.4 and 6.5 can also be compared with the relatively high number of users who admitted they rarely or never contribute because they do not know it is possible or do not know how to configure the Contribute functionality (seven out of 28 who rarely or never contribute).

Among the other comments, some users also proposed to group free journals by domains (such as.gov) or by countries.

Three respondents from the United States suggested in Question 2 and Question 6 to create specific targets by language. According to them, grouping journals by language (at least for the main ones) in separate targets would probably be helpful to SFX administrators. Moreover, it would certainly be easier to create than a target based on the scholarly nature of the titles. However, this proposal has not been submitted to the SFX community through the survey, and making such separate targets also relies heavily on the metadata quality of the objects themselves.

### Question 7: What Kinds of Particular Problems Do You Have With This Target? Please Try and Classify Them From Largest to Least Important Problem

The dissatisfaction with this target expressed by many respondents is a reaction to what may be summed up as its unreliability, as evidenced in a number of specific problems. Of the 114 users of the target, 84 respondents (73.7%) have left a comment on this open question. Some answers only focus on one or two problems, while others point out up to five. Three issues come up clearly and are very often classified in the top level:

(1) Incorrect parse params, broken links—mentioned by 52 respondents (61.9%).
(2) Full texts actually not free (subscription required, hybrid journal)—mentioned by 35 respondents (41.7%).
(3) Incorrect or missing thresholds—mentioned by 32 respondents (38.1%).

Among the other problems, eight have been mentioned by at least two respondents (see Figure 12):

(1) Journal also available in other OA or free targets (9).
(2) Linking only available at journal level, not article (5).
(3) Target is too big, making it difficult to maintain by Ex Libris and the clients (4).
(4) Too many languages (4).
(5) Too many mixed subjects, disciplines (4).
(6) Poor quality of the KB object metadata (4).
(7) Nonacademic journals (3).
(8) Duplicates within the target (2).

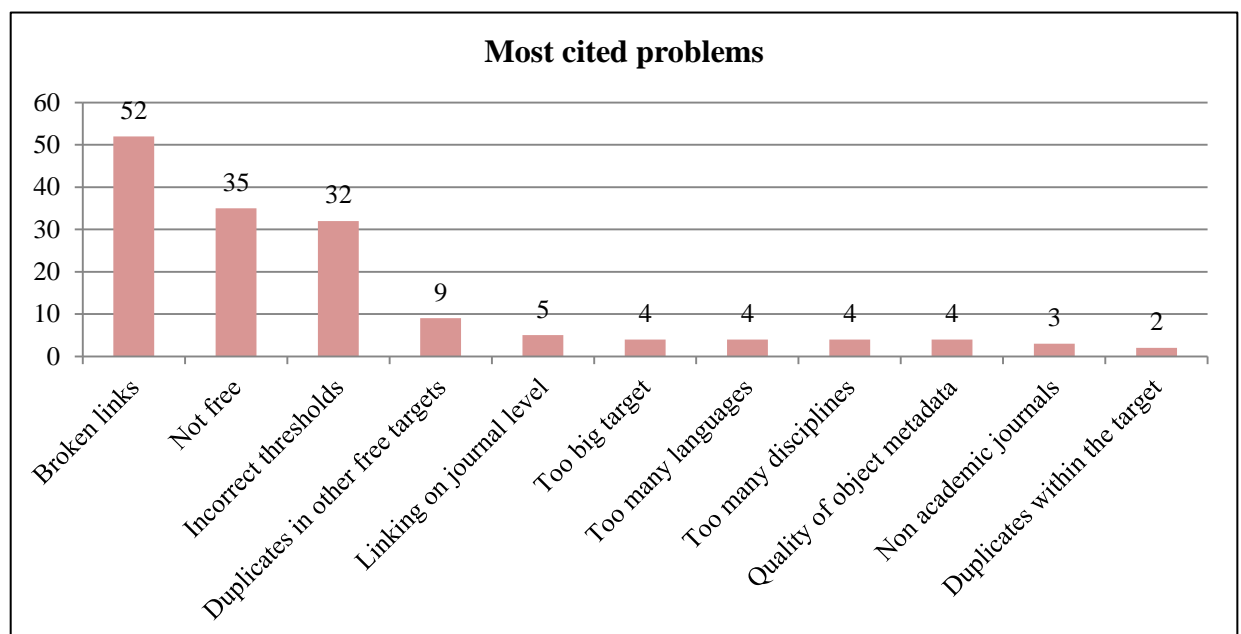

*Figure 12. Most-cited problems.*

There is a general view that the target is not maintained well enough by Ex Libris and perhaps is not a top priority, leading to the problems experienced—in addition to problems that are the responsibility of the journals' publishers.

One respondent commented: "With users being accustomed to use of Google and Google Scholar (often as first choice source) they may find free material on their own. Is this target atavistic?" Another says the target as it is gives satisfaction: "I am glad this target exists and I do use it. I do not think anything needs to change."

One suggestion for dealing with some of the target's problems (too big to be adequately maintained) is to break it up by creating specific OA targets for topics or individual publishers, e.g., OXFORD_OPEN, THIEME_OPEN.

One respondent reported having "found more quality control problems since users can Contribute, e.g., URL doesn't work on a newly added OP [Object Portfolio]." The broken-link problem for newly added portfolios has also been pointed out by another user.

Finally, some respondents made software suggestions that could improve the target usage:

- Used sources: "I do not know what the source is for these free portfolios but I feel like once a portfolio is created there isn't any follow up. It's nice to be provided with free options, but if they are going to be provided they should be monitored and updated. When this survey is complete and the results hopefully shared, maybe you could also include the 'source/s' for these portfolios if you know."
- Possibility of integration with other systems (i.e., a process for requesting ISSNs, to create CONSER cataloging records).

One of the most interesting developments that has been occurring in recent years and one that will have great impact on link resolvers like SFX is the growth of hybrid journals. Hybrid journals contain both licensed and OA content. Since the SFX KB only maintains information at the journal level (and not at the article level), being able to provide free access to the OA articles in a hybrid journal while maintaining authenticated access to the other material in the journal poses some interesting challenges that will need to be dealt with, especially as the number of hybrid journals continues to increase.

A related challenge is the desire of link resolver customers to have the product provide access to other content types besides the full text of journal articles with which they currently deal. Resolvers deal with ebooks (although they pose some interesting problems), but customers are beginning to want their resolvers to provide access to other content type like audio, video, multimedia, and the like. The OpenURL framework introduced by the NISO standard provides the mechanism for describing such content in an OpenURL. However, how access to that content will be made available in link resolvers like SFX will be an interesting challenge.

## Question 8: Other Comments You Might Have About MISCELLANEOUS_FREE_EJOURNALS

This last question was open to every respondent; even those who answered "No" to the first question could leave a final comment. Most of the 30 comments echo issues raised in Question 7 regarding the target:

- Unreliability.
- Redundant portfolios (with other free targets).
- Difficult to manage.
- Poor maintenance and questionable commitment on the part of Ex Libris.
- Too large.
- Lack of clarity regarding criteria for inclusion.
- Suspicions regarding some OA/free titles.
- Patrons are confused and inconvenienced by problematic aspects of the target, requiring explanations and caution from the library, such as disclaimers in the user interface.

At the same time, it is seen as potentially valuable if efforts could be made to address the drawbacks and progressively correct some of them. MFE has the merits of introducing users to many journals that were hitherto unknown to them and making accessible journals in local languages not covered by other targets. It seems that precisely because of the collection's great potential, there is frustration and disappointment due to its various problems. In a sense, the strength of MFE is also its weakness.

**CONCLUSIONS**

Due to the dynamic nature of online journals and many other factors, maintaining the KB for SFX or any other OpenURL link resolver has been and will be a challenge. The SFX MFE target is particularly challenging, as found by the SFX Knowledge Base Advisory Board (KBAB), with the error rate in parse params as high as 20%. On the other hand, as the number of free or OA journals in the world is exploding, information professionals and the information industry cannot afford to ignore them.

The survey done by KBAB in 2014 offered an overview of how the SFX MFE target is used by libraries across the world. The survey found that 93.4% respondents use the target, but most (57.9%) use it selectively, and 50.9% do not use the SFX "auto-activate" function for this target, meaning that these libraries do not want to automatically accept journals added to this collection by SFX. As evidence that most librarians not only use this free journals target on SFX but also actively make contributions to the target, 68.4% respondents have added free journals (at least locally), and 90.4% corrected errors in coverage years and links in this target in one way or another. A total of 83.3% respondents think that Ex Libris should check the portfolios of this target more systematically. The majority of respondents like the ideas of reorganizing and rearranging the journals in the target by scholarly/academic nature and by "de-duplicating" the target from other free journal targets. A few respondents wrote suggestions that journals in this target be regrouped by language.

When asked to list the biggest problems of this target, respondents gave these top three problems: (a) incorrect parse params (broken links); (b) full texts actually not free (subscription required, hybrid journal); and (c) incorrect or missing thresholds. In summary, libraries would like to use this target but found that it has more problems than other targets.

Some potential solutions were also discussed by the respondents of this survey. We hope Ex Libris will take the respondents' comments and suggestions into account in order to improve this widely used target whose quality relies exclusively on the Ex Libris KB team's work and the collaboration of the clients. Other OpenURL link resolver vendors and users may find the survey and suggestions relevant too, if they offer links to free or OA journals. Various changes could be made to strengthen the SFX MFE target. De-duplicating the target from other existing free or OA targets, especially DOAJ, would reduce the size of the target and ease the burden of managing OA journals already included in DOAJ and other OA targets. After all, all other OA targets have a vendor or owner who manages the contents. Grouping journals by language in separate targets would be another way to reduce the size of the existing target and would also make the work of SFX administrators easier in selecting and managing targets for their specific users' needs. Finally, checking the portfolios of this target (whether or not this target is regrouped) more systematically is the ultimate quality control Ex Libris should implement, and other changes such as de-duplicating, de-selecting, and regrouping would make it easier to implement this important quality control.

**APPENDIX**

The survey was composed of eight questions related to the MISCELLANEOUS_FREE_EJOURNALS target:

1. Do you use the MISCELLANEOUS _FREE_ EJOURNALS target?
   - Yes
   - No

If not, please explain why not.

2. Do you **activate all of the object portfolios** in the target or only selectively?
   - All portfolios
   - Only a selection

Please explain why.

3. Do you set **auto-activate** on for this target?
   - Yes
   - No

If not, why not?

4. Do you **add local** object portfolios to this target?
   - Yes, often
   - Yes, sometimes
   - Rarely
   - Never

If yes, do you contribute? Do you request Ex Libris to **add** them to the KB?
   - Yes, often
   - Yes, sometimes
   - Rarely
   - Never

5. Do you **locally correct** thresholds or parse params of any of the portfolios in this target?
   - Yes, often
   - Yes, sometimes
   - Rarely
   - Never

If yes, do you contribute? Do you request Ex Libris to **correct** them in the KB?
If rarely or never, please explain why.

6. What would you like to see happen with this target? What could Ex Libris do, in your opinion, **to improve this target?**
Legend: 1 = *Least Desirable;* 5 = *Very Desirable*
   - Checking the portfolios (thresholds and parse params) more systematically?
   - Creating a new specific target MISCELLANEOUS. FREE_SCHOLARLY_EJOURNALS for academic journals only? All the other journals would stay in MISCELLANEOUS_FREE_EJOURNALS.

- Deleting from MISCELLANEOUS_FREE_ EJOURNALS all portfolios that already exist in another FREE or OPEN-ACCESS target?
- Promoting the use of Contribute button?
- Getting a pop-up invitation to use the Contribute button every time I have locally updated a threshold or a parse param?

7. What kinds of **particular problems** do you have with this target? Please try and classify them from largest to least important problem.

8. Other comments you might have about MISCELLANEOUS _FREE_EJOURNALS.